\documentclass[11pt]{article}

\usepackage{epsfig,subfigure}
\usepackage{amssymb, amsmath}
\usepackage{color}
\usepackage{hyperref}
\usepackage{graphicx}
\usepackage{color}

\usepackage{amssymb}
\usepackage{amsmath,amsfonts,amssymb}

\usepackage{verbatim}
\usepackage{stfloats}
\usepackage[bookmarks=false]{}

\newtheorem{theorem}{Theorem}

\newtheorem{definition}{Definition}
\newtheorem{lemma}{Lemma}

\newcommand\blfootnote[1]{%
	\begingroup
	\renewcommand\thefootnote{}\footnote{#1}%
	\addtocounter{footnote}{-1}%
	\endgroup
}

\begin{document}
\date{}
\title{Aligned Image Sets and the {\color{black}Generalized Degrees of Freedom} of  Symmetric MIMO Interference Channel with  Partial CSIT}
\author{ \normalsize Arash Gholami Davoodi and Syed A. Jafar \\
{\small Center for Pervasive Communications and Computing (CPCC)}\\
{\small University of California Irvine, Irvine, CA 92697}\\
{\small \it Email: \{gholamid, syed\}@uci.edu}
}
\maketitle

\blfootnote{This work will be presented in part at IEEE GLOBECOM 2017.}

\begin{abstract}

The generalized degrees of freedom (GDoF) of the two user symmetric multiple input multiple output (MIMO) interference channel (IC) are characterized as a function of the channel strength levels and the level of channel state information at the transmitters (CSIT). In this symmetric setting, each transmitter is equipped with $M$ antennas, each receiver is equipped with $N$  antennas, and both cross links have the same strength parameter $\alpha$ and the same channel uncertainty parameter $\beta$. The main challenge resides in the proof of the outer bound which is accomplished by a generalization of the aligned image sets approach. 

\end{abstract}
\section{Introduction}
The pursuit of progressively refined capacity approximations over the past decade has  produced numerous new insights  into the fundamental limits of wireless networks. While degrees of freedom (DoF) studies are often the starting point, a GDoF characterization is the natural next step forward along this path. It is also a most significant step forward, because unlike the DoF metric which is not capable of making distinctions based on channel strength levels (any non-zero channel carries $1$ DoF) or partial CSIT levels (finite precision CSIT is equivalent to no CSIT, both cause collapse of DoF \cite{Arash_Jafar_PN}), GDoF is sensitive to both channel strengths and channel uncertainty levels. As such, GDoF characterizations are capable of shedding light on  optimal yet \emph{robust} interference management schemes for settings where interference may be significantly weaker or stronger than desired signals, and where the channel state information at the transmitters (CSIT) is neither perfect nor so weak as to be ignored entirely.

A critical barrier for GDoF characterizations, especially under partial CSIT, has been the difficulty of obtaining tight outer bounds for these settings. Notably, the 2005 conjecture of Lapidoth et al. in \cite{Lapidoth_Shamai_Wigger_BC}, which claimed that the DoF of wireless networks should collapse under finite precision CSIT, was only settled recently in \cite{Arash_Jafar_PN} by introducing a novel aligned image sets (AIS) approach. The original argument of \cite{Arash_Jafar_PN} is based on a combinatorial accounting of the size of the aligned image sets under finite precision channel knowledge. Several recent works have successfully built upon the AIS argument to obtain new GDoF characterizations. The GDoF of the $2$ user MISO BC are characterized in \cite{Arash_Jafar_TC} for arbitrary channel strength levels and arbitrary channel uncertainty levels for each channel coefficient. The GDoF are obtained for the $K$ user symmetric IC under finite precision CSIT in  \cite{Arash_Jafar_IC}, and for symmetric instances of $K$ user MISO BC  in \cite{Arash_Bofeng_Jafar_BC}.  Most recently, in \cite{Arash_Jafar_sumset}, the AIS approach is further generalized to present sum-set inequalities specialized to the GDoF framework. Building upon these recent advances, in this work we explore the GDoF  of the two user MIMO interference channel (IC). 

For the MIMO IC previous works have explored the impact of different channel strengths through DoF and GDoF characterizations under perfect CSIT \cite{Jafar_Fakhereddin, Karmakar_Varanasi}. The impact of limited CSIT is explored through DoF characterizations under no CSIT \cite{Huang_Jafar_Shamai_Vishwanath, Zhu_Guo_MIMOIC, Varanasi_noCSIT}. Most recently, the DoF region of the MIMO IC under partial CSIT with arbitrary antenna configurations is settled in \cite{Bofeng_Arash_Jafar_ArXiv} based on the sum-set inequalities of  \cite{Arash_Jafar_sumset}. As the next step, in this work we explore the \emph{joint} impact of channel strength levels and partial channel knowledge for the two user MIMO IC. To this end, we characterize the GDoF of the symmetric MIMO IC, where each transmitter is equipped with $M$ antennas, each receiver is equipped with $N$ antennas, and where each cross-channel has channel strength parameter $\alpha$ and CSIT level $\beta$, for arbitrary values of $M, N, \alpha,\beta$. While the restrictive assumptions of symmetry are enforced to avoid an explosion in the number of parameters, the key ideas from this work should generalize to asymmetric settings as well. Notably, this is the first application of the AIS argument to jointly deal with multiple spatial dimensions at both transmitters and receivers, in conjunction with different channel strengths and partial CSIT levels.


{\it Notation:} For $n\in\mathbb{N}$, define the notation $[n]=\{1,2,\cdots,n\}$. The cardinality of a set $A$ is denoted as $|A|$. The notation $X_{1:i}$ stands for $\{X_1, X_2, \cdots, X_i\}$ and $X^{[n]}$ stands for $X(1), X(2), \cdots X(n)$. Moreover,  $X_{1:k}^{[n]}$ also stands for $\{X_i(t): \forall i\in [k], \forall t\in[n]\}$. For sets $A, B$, the notation $A/B$ refers to the set of elements that are in $A$ but not in $B$. Moreover, we use the Landau $O(\cdot)$, $o(\cdot)$, and $\Theta(\cdot)$ notations as follows. For  functions $f(x), g(x)$ from $\mathbb{R}$ to $\mathbb{R}$, $f(x)=O(g(x))$ denotes that $\limsup_{x\rightarrow\infty}\frac{|f(x)|}{|g(x)|}<\infty$.  $f(x)=o(g(x))$ denotes that $\limsup_{x\rightarrow\infty}\frac{|f(x)|}{|g(x)|}=0$. $f(x)=\Theta(g(x))$ denotes that there exists a positive finite constant, $M$,  such that $\frac{1}{M} g(x)\leq f(x)\leq Mg(x)$, $\forall x$. We use $\mathbb{P}(\cdot)$ to denote the probability function $\mbox{Prob}(\cdot)$. We define $\lfloor x\rfloor$ as the largest integer that is smaller than or equal to $x$ when $x>0$,  the smallest integer that is larger than or equal to $x$ when $x<0$, and $x$ itself when $x$ is an integer. 

\section{Definitions}
\begin{definition}[Bounded Density Channel Coefficients] \label{deffp}Define a set of real-valued random variables, $\mathcal{G}$ such that the magnitude of each random variable $g\in\mathcal{G}$ is bounded away from infinity, $ |g|\leq\Delta_2<\infty$, for some positive constant $\Delta_2\geq 1$, and there exists a finite positive constant $f_{\max}\geq 1$, such that for all finite cardinality disjoint subsets $\mathcal{G}_1, \mathcal{G}_2$ of $\mathcal{G}$, the joint probability density function of all random variables in $\mathcal{G}_1$, conditioned on all random variables in $\mathcal{G}_2$, exists and is bounded above by $f_{\max}^{|\mathcal{G}_1|}$. \end{definition}

\begin{definition}[Arbitrary Channel Coefficients] \label{defarbit} Let $\mathcal{H}$ be a set of arbitrary constant values that are  bounded above by $\Delta_2$, i.e., if $h\in\mathcal{H}$ then $|h|\leq\Delta_2<\infty$. 
\end{definition}

\begin{definition}\label{ss} For any positive number $\alpha_i$, define alphabet $\mathcal{X}_{\alpha_i}$ as,
\begin{eqnarray}
\mathcal{X}_{\alpha_i}&\triangleq&\{0,1,2,\cdots,\bar{P}^{\alpha_i}\}
\end{eqnarray}
where $\bar{P}^{\alpha_i}$ is a compact notation for $\left\lfloor\sqrt{P^{\alpha_i}}\right\rfloor$. For $X\in\mathcal{X}_\alpha$, and $0\leq \alpha'\leq\alpha$, define \begin{eqnarray}
(X)^{\alpha'}&\triangleq&\left \lfloor \frac{X}{\bar{P}^{\alpha-\alpha'}} \right \rfloor
\end{eqnarray}
\end {definition}
In words, $(X)^{\alpha'}$ retrieves the top $\alpha'$ power levels of $X$.
\begin {definition} ~For~ real~ numbers~ $v_1,v_2,\cdots,v_k$ ~and the vector $V=\begin{bmatrix}v_1&v_2&\cdots&v_k\end{bmatrix}^T$ define the notations $L^b_j(V)$ and $L_j(V)$ to represent,
\begin {align}
L^b_j(V)\triangleq&\sum_{1\le i\le k} \lfloor g_{j_i}v_i\rfloor\\
L_j(V)\triangleq&\sum_{1\le i\le k} \lfloor h_{j_i}v_i\rfloor
\end{align}
for  distinct random variables $g_{j_i}\in\mathcal{G}$ and $h_{j_i}\in\mathcal{H}$. We refer to the $L^b$ functions as the bounded density linear combinations.
\end {definition}
\begin{definition}\label{defvec} For any vector $V=\begin{bmatrix}v_1&\cdots&v_k\end{bmatrix}^T$ and non-negative integer numbers $m$ and $n$ less than $k$, define 
\begin{eqnarray}
V_{m,n}&\triangleq&\left\{
\begin{array}{ll}
\begin{bmatrix}v_{m+1}&\cdots&v_{m+n}\end{bmatrix}^T,& m+n\le k\\
\begin{bmatrix}v_{m+1}&\cdots&v_k&v_1&\cdots&v_{m+n-k}\end{bmatrix}^T,& k<m+n
\end{array}
\right.
\end{eqnarray}
Moreover, for the two vectors $V=\begin{bmatrix}v_1&\cdots&v_{k_1}\end{bmatrix}^T$ and $W=\begin{bmatrix}w_1&\cdots&w_{k_2}\end{bmatrix}^T$ define $V;W$ as $\begin{bmatrix}v_1&\cdots&v_{k_1}&w_1&\cdots&w_{k_2}\end{bmatrix}^T$.
\end{definition}

\section{System Model} {\label{sec-sys}}
For ease of exposition, in this work we will focus on the setting where all variables take only real values. Extensions to complex settings are cumbersome but conceptually straightforward as shown in  \cite{Arash_Jafar_PN}.
\subsection{The Channel}
\begin{figure}[tp]
\centering 
\includegraphics[scale =0.4]{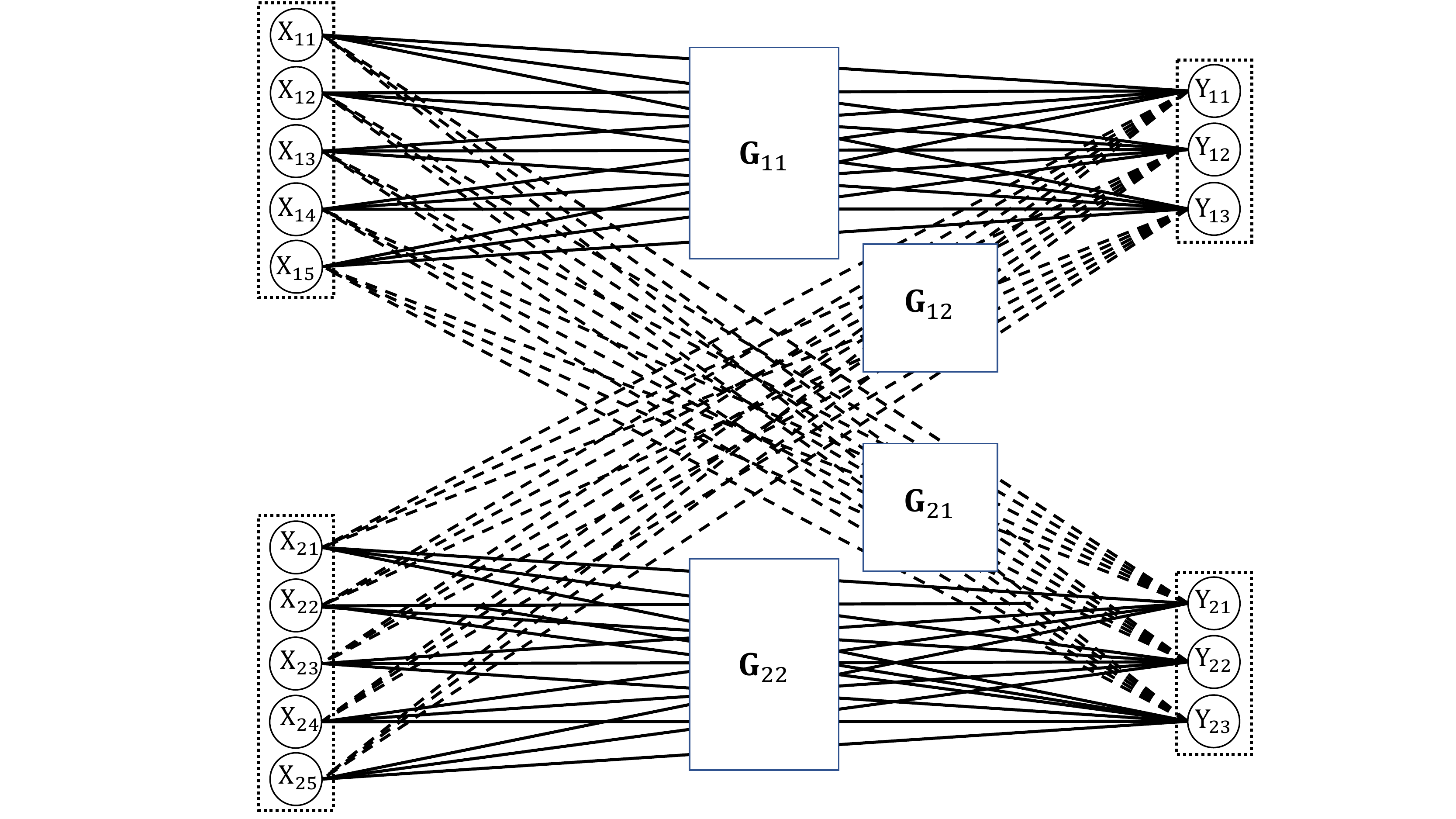}
\caption{Two user $5 \times 3$ MIMO IC.}
\label{Fig1}
\end{figure}

Define the random variables $\mathbf{X}_{s}(t)$ and $\mathbf{Y}_{r}(t)$ for $r,s\in\{1,2\}$ as,
\begin{align}
\mathbf{X}_{s}(t)=&\begin{bmatrix}{X}_{s1}(t)&{X}_{s2}(t)&\cdots&{X}_{sM}(t)\end{bmatrix}^T\\
\mathbf{Y}_{r}(t)=&\begin{bmatrix}{Y}_{r1}(t)&{Y}_{r2}(t)&\cdots&{Y}_{rN}(t)\end{bmatrix}^T
\end{align}
The channel uses are indexed by $t\in\mathbb{N}$, $X_{sm}(t), s\in\{1,2\},m\in[M]$ are the symbols sent from $m$-th transmit antenna of the $s$-th transmitter and are subject to unit power constraint, while $Y_{rn}(t), r\in\{1,2\},n\in[N]$ are the symbols observed by the $n$-th antenna of the $r$-th receiver. Under the GDoF framework, the channel model for the two user MIMO IC is defined by the following input-output equations
\begin{align}
\mathbf{Y}_{i}(t)=&\sqrt{P}{\bf G}_{ii}(t)\mathbf{X}_{i}(t)+\sqrt{P^{\alpha}}{\bf G}_{i\bar{i}}(t)\mathbf{X}_{\bar{i}}(t)+\mathbf{\Gamma}_{i}(t),~\forall i\in\{1,2\}\label{TR21}
\end{align}
Here we have defined $\bar{i}=3-i$, so that $\bar{i}=1$ if $i=2$ and $\bar{i}=2$ if $i=1$.
The $N\times M$ matrix ${\bf G}_{rs}(t)$ is the channel fading coefficient matrix between the $r$-th receiver and the $s$-th transmitter for any $r,s\in\{1,2\}$. 
The entry in the $n$-th row and $m$-th column of the matrix ${\bf G}_{rs}(t)$ is ${G}_{rsnm}(t)$. $\mathbf{\Gamma}_{1}(t)$ and $\mathbf{\Gamma}_{2}(t)$ are $N\times 1$  matrices whose components are zero mean unit variance additive white Gaussian noise (AWGN). See Fig \ref{Fig1} for two user $5\times3$ MIMO IC. $P$ is the nominal $SNR$ parameter that approaches infinity for the GDoF characterizations. Channel state information at the receivers (CSIR) is assumed to be perfect. However, the channel state information at the transmitters (CSIT) is only partially available, as specified next. 

\subsubsection{Partial CSIT} \label{defp}
Under partial CSIT, the channel coefficients are represented as
\begin{eqnarray*}
G_{rsnm}(t)&=&\hat{G}_{rsnm}(t)+\sqrt{P^{-\beta_{rs}}}\tilde{G}_{rsnm}(t)
\end{eqnarray*}
Recall that $G_{rsnm}(t)$ is the channel fading coefficient between the $n$-th antenna of $r$-th receiver and $m$-th antenna of  $s$-th transmitter. $\hat{G}_{rsnm}(t)$ is the  channel estimate and $\tilde{G}_{rsnm}(t)$ is the  estimation error term. To avoid degenerate conditions, for each $N\times M$ channel matrix ${\bf G}_{rs}(t)$, we require that all its $N\times N$ submatrices are non-singular, i.e., their determinants are bound away from zero. To this end, if $N\le M$, then for all $t\in[n],~r,s\in\{1,2\}$, and for all choices of $N$ transmit antenna indices $\{m_1,m_2,\cdots,m_N:m_i\in [M]\}$ define the determinant $D(t)$ as
\begin{eqnarray}
D(t)\triangleq \begin{vmatrix}
 G_{rs1m_1}(t)& G_{rs1m_2}(t)& \cdots & G_{rs1m_N}(t)\\ 
 \vdots&\vdots  & \ddots  &\vdots \\ 
 G_{rsNm_1}(t)& G_{rsNm_2}(t)&  \cdots& G_{rsNm_N}(t)
\end{vmatrix}.\label{deter}
\end{eqnarray}
Then we require that there exists a positive constant $\Delta_1>0$, such that $|D(t)|\geq \Delta_1$, for all $t\in[n],~r,s\in\{1,2\},\{m_1,m_2,\cdots,m_N:m_i\in [M]\}.$
The channel variables $\hat{G}_{rsnm}(t), \tilde{G}_{rsnm}(t)$ are distinct random variables drawn from the set $\mathcal{G}$. The  realizations of $\hat{G}_{rsnm}(t)$ are known to the transmitter, but the realizations of $\tilde{G}_{rsnm}(t)$ are not available to the transmitter. We also assume that the channel coefficients $|{G}_{rsnm}(t)|$ are bounded away from zero, i.e., 
\begin{align}
\Delta_1\le|{G}_{rsnm}(t)|, \forall t\in[n],~r,s\in\{1,2\},m\in [M],n\in [N]\label{Deter2}
\end{align}
 Note that under the partial CSIT model, the variance of the channel coefficients $G_{rsnm}(t)$  behaves as $\sim P^{-\beta_{rs}}$ and the peak of the probability density function  behaves as $\sim\sqrt{P^{\beta_{rs}}}$.

For any $r,s\in\{1,2\}$, in order to span the full range of partial channel knowledge at the transmitters, the corresponding range of $\beta_{rs}$ parameters, assumed throughout this work, is $0\leq\beta_{rs}\leq\alpha$. $\beta_{rs}=0$ and $\beta_{rs}=\alpha$ correspond to the two extremes where the CSIT is essentially absent, or perfect, respectively. Note that the value of $\beta_{11}$ and $\beta_{22}$ will not affect the GDoF.

\subsection{GDoF}
The definitions of achievable rates $R_i(P)$ and capacity region $\mathcal{C}(P)$ are standard. The GDoF region is defined as
\begin{align}
\mathcal{D}=&\{(d_1,d_2): \exists (R_1(P),R_2(P))\in\mathcal{C}(P), \nonumber\\
&\mbox{ s.t. } d_k=\lim_{P\rightarrow\infty}\frac{R_k(P)}{\frac{1}{2}\log{(P)}}, \forall k\in\{1,2\}\} \label {region}
\end{align}

\section{Main Result}
For $M\leq N$, the GDoF of the MIMO IC with partial CSIT are the same as with perfect CSIT for which the result is already known \cite{Karmakar_Varanasi}.  So, henceforth, $N< M$ is assumed throughout this paper.

\begin{theorem}\label{theorem:GDoF} The sum GDoF value for the two user symmetric MIMO IC for $N< M$ is,
\begin{align}
&d_1+d_2=
&\left\{
\begin{array}{ll} 
2N(1-\alpha)+2\hat{N}\beta,&0\leq\alpha\leq \frac{1}{2}\vspace{0.03in}\\ 
2N\alpha+2\hat{N}(1-2\alpha+\beta)^+, &\frac{1}{2}<\alpha\leq\frac{2}{3}\vspace{0.03in}\\ 
\min(2N\alpha+2\hat{N}(1-2\alpha+\beta)^+,2N-N\alpha+\hat{N}\beta),&\frac{2}{3}<\alpha\leq 1\\
\min(2N,N\alpha+\hat{N}(1-\alpha+\beta)^+),& 1<\alpha\leq 2\\
2N, &2<\alpha
\end{array}
\right.\label {GDOFM>N}
\end{align}
where $\beta\le\alpha$ and $\hat{N}$ is defined as $\min(N,M-N)$. Note that the sum GDoF value for $\beta>\alpha$ is the same as with perfect CSIT, i.e., $\beta=\alpha$.
\end{theorem}

\section{Proof of Theorem \ref{theorem:GDoF}: Converse}
\subsection{Equivalent Channel for Outer Bound}\label{CoB_1}
\begin{figure}[tp]
\centering 
\includegraphics[scale =0.4]{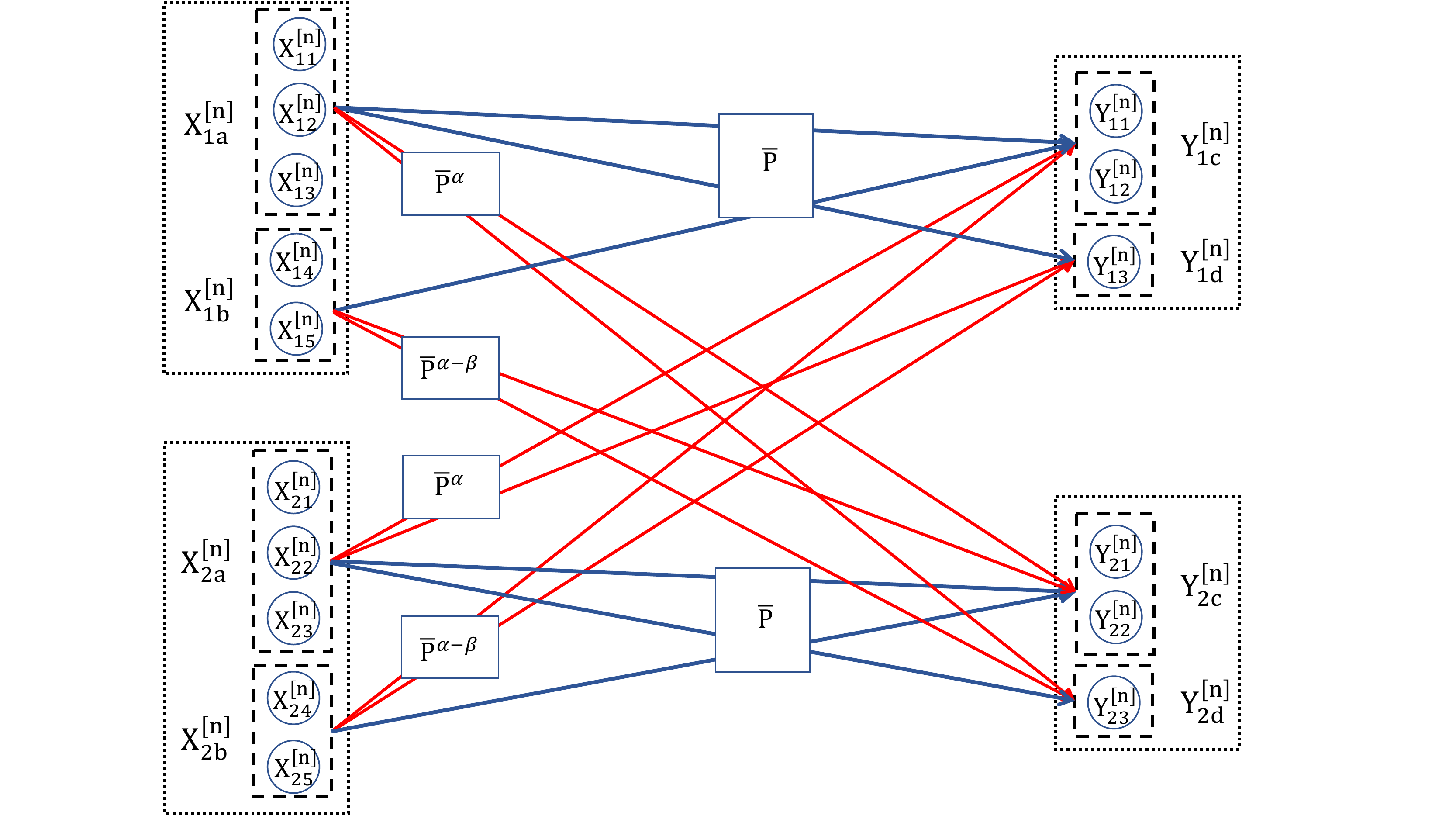}
\caption{Equivalent channel for two user $5 \times 3$ MIMO IC. The channel strength terms, e.g., $\bar{P}^{\alpha-\beta}$ represent the maximum \emph{received} signal strength that can be delivered by the corresponding channels.}
\label{Fig2}
\end{figure}

Without loss of generality, we can perform a sequence of invertible operations (specifically, multiplications of inputs and outputs by unitary matrices) that are inconsequential for GDoF, similar to \cite{Gou_Jafar}, at the transmitters and receivers to convert the channel to a simpler form. For instance, the equivalent channel for a $5\times3$ MIMO interference channel is depicted in Fig \ref{Fig2}.

In the equivalent channel, the transmitted symbol vector at time $t$ for transmitter $s\in\{1,2\}$, $\mathbf{X}_s' (t)$ of size $M\times 1$, is partitioned  into $\mathbf{X}'_{sa}(t)$ and $\mathbf{X}'_{sb}(t)$ as,
\begin{align}
\mathbf{X}'_{sa}(t)=&{[\mathbf{X}'_{s}(t)]}_{0,N}\label{vve1}\\
\mathbf{X}'_{sb}(t)=&{[\mathbf{X}'_{s}(t)]}_{N,M-N}\label{vve2}
\end{align}
For any $i\in\{1,2\}$, the matrix $\hat{\mathbf{G}}_{\bar{i}i}(t)$ has $M-N$ null space dimensions. So the $i$-th transmitter can  zero-force $\mathbf{X}'_{{i}b}(t)$ into the null space of $\hat{\mathbf{G}}_{\bar{i}i}(t)$ in a way that the $\bar{i}$-th receiver sees only the top $\alpha-\beta$ power levels of $\mathbf{X}'_{{i}b}(t)$. Define $\breve{N}=(2N-M)^+$. Note that $N=\breve{N}+\hat{N}$. In the equivalent channel, the $N\times 1$ output signal vector at receiver $i$,  $\mathbf{Y}'_{i}(t)$ is partitioned into two $\hat{N}\times 1$ and $\breve{N}\times 1$ vectors, $\mathbf{Y}'_{ic}(t)$ and $\mathbf{Y}'_{id}(t)$, i.e.,
\begin{align}
\mathbf{Y}'_{ic}(t)=&\sqrt{P}{\bf G}'_{iia}(t)\mathbf{X}'_{ia}(t)+\sqrt{P}{\bf G}''_{iia}(t)\mathbf{X}'_{ib}(t)+\sqrt{P^{\alpha}}{\bf G}'_{i\bar{i}a}(t)\mathbf{X}'_{\bar{i}a}(t)\nonumber\\
&+\sqrt{P^{\alpha-\beta}}\tilde{\bf G}''_{i\bar{i}a}(t)\mathbf{X}'_{\bar{i}b}(t)+\mathbf{\Gamma}'_{i}(t),~~\forall i\in\{1,2\},t\in[n]\label{re1}\\
\mathbf{Y}'_{id}(t)=&\sqrt{P}{\bf G}'_{iib}(t)\mathbf{X}'_{ia}(t)+\sqrt{P^{\alpha}}{\bf G}'_{i\bar{i}b}(t)\mathbf{X}'_{\bar{i}a}(t)\nonumber\\
&+\sqrt{P^{\alpha-\beta}}\tilde{\bf G}''_{i\bar{i}b}(t)\mathbf{X}'_{\bar{i}b}(t)+\mathbf{\Gamma}''_{i}(t),~~\forall i\in\{1,2\},t\in[n]\label{re2}
\end{align}
where for any $i\in\{1,2\}$, $\mathbf{X}_{ib}(t)$ does not appear at $\mathbf{Y}_{id}(t)$. Moreover, for any $i\in\{1,2\}$, $t\in[n]$, ${\bf G}'_{iia}(t)$, ${\bf G}''_{iia}(t)$, ${\bf G}'_{i\bar{i}a}(t)$ and $\tilde{\bf G}''_{i\bar{i}a}(t)$ are $\hat{N}\times N$, $\hat{N}\times (M-N)$, $\hat{N}\times N$ and $\hat{N}\times (M-N)$ matrices while ${\bf G}'_{iib}(t)$, ${\bf G}'_{i\bar{i}b}(t)$ and $\tilde{\bf G}''_{i\bar{i}b}(t)$  are $\breve{N}\times N$, $\breve{N}\times N$ and $\breve{N}\times (M-N)$ matrices respectively.\footnote{For any $i\in\{1,2\}$ consider an invertible $M\times M$ matrix ${\bf U}_i(t)$ with unit determinant where $\hat{\bf G}_{\bar{i}i}(t){\bf U}_i(t)$'s right $M-N$ columns are zero. Note that this is possible because the matrix $\hat{\mathbf{G}}_{\bar{i}i}(t)$ has $M-N$ null space dimensions. So, we perform an invertible linear transformation at the transmitters by multiplying ${\bf U}_i(t)$ to the transmitted signal at the $i$-th transmitter, i.e., $\mathbf{X}_{i}(t)={\bf U}_i(t)\mathbf{X}'_{i}(t)$ and transmit $\mathbf{X}'_{i}(t)$ instead of $\mathbf{X}_{i}(t)$. Moreover, consider an invertible $N\times N$ matrix ${\bf U}'_i(t)$ with unit determinant such that ${\bf U}'_i(t){\bf G}_{ii}{\bf U}_i(t)(t)$'s lower right $\breve{N}\times (M-N)$ block is the zero matrix. From (\ref{TR21}) we have,
\begin{align}
\mathbf{Y}'_{i}(t)=&\sqrt{P}{\bf U}'_i(t){\bf G}_{ii}(t){\bf U}_i(t)\mathbf{X}'_{i}(t)+\sqrt{P^{\alpha}}{\bf U}'_i(t){\bf G}_{i\bar{i}}(t){\bf U}_{\bar{i}}(t)\mathbf{X}'_{\bar{i}}(t)+{\bf U}'_i(t)\mathbf{\Gamma}_{i}(t),~\forall i\in\{1,2\}\label{ffgf}
\end{align}
from (\ref{ffgf}), the equivalent channel (\ref{re1}) and (\ref{re2}) are concluded. } ${\bf \Gamma}'_{i}(t)$ and ${\bf \Gamma}''_{i}(t)$ are also  $\hat{N}\times 1$  and $\breve{N}\times 1$ matrices whose components are zero mean unit variance AWGN. Note that because the equivalent channel is obtained by simply rotating the input and output vectors (multiplications by unitary matrices) at each transmitter and receiver, all the transmit power constraints and the assumptions on the channel coefficients specified in Section \ref{sec-sys} are inherited by the equivalent channel as well.

\subsection{Deterministic Model}\label{DM_1}
As in \cite{Arash_Jafar_IC}, without loss of generality for GDoF characterizations, we will use the deterministic model for the equivalent channel.
\begin{align}
\bar{\mathbf{Y}}_{ic}(t)&=L_{i1}(t)\left({(\bar{\mathbf{X}}_{ia}(t))}^1;{(\bar{\mathbf{X}}_{ib}(t))}^1;{(\bar{\mathbf{X}}_{\bar{i}a}(t))}^{\alpha}\right)+L_{i1}^b(t)\left({(\bar{\mathbf{X}}_{\bar{i}b}(t))}^{\alpha-\beta}\right)\label{dm1}\\
\bar{\mathbf{Y}}_{id}(t)&=L_{i2}(t)\left({(\bar{\mathbf{X}}_{ia}(t))}^1;{(\bar{\mathbf{X}}_{\bar{i}a}(t))}^{\alpha}\right)+L_{i2}^b(t)\left({(\bar{\mathbf{X}}_{\bar{i}b}(t))}^{\alpha-\beta}\right)\label{dm2}
\end{align}
where $\bar{\mathbf{X}}_{i}(t)$, $\bar{\mathbf{Y}}_{ic}(t)$ and $\bar{\mathbf{Y}}_{id}(t)$ are integer-valued vectors. $\bar{\mathbf{X}}_{i}(t)$, $\bar{\mathbf{X}}_{ia}(t)$ and $\bar{\mathbf{X}}_{ib}(t)$ are defined from (\ref{vve1}) and (\ref{vve2}) as,
\begin{align}
\bar{\mathbf{X}}_{i}(t)=&[\bar{X}_{i1}(t)\ \bar{X}_{i2}(t)\ \cdots\ \bar{X}_{iM}(t)]^T\label{ggf1}\\
\bar{\mathbf{X}}_{ia}(t)=&{[\bar{\mathbf{X}}_{i}(t)]}_{0,N}\\
\bar{\mathbf{X}}_{ib}(t)=&{[\bar{\mathbf{X}}_{i}(t)]}_{N,M-N}\label{ggf2}
\end{align}
and $\bar{X}_{im}(t)\in\{0, 1, \cdots, {\bar{P}}^{\max(1,\alpha)}\}$, $\forall m\in[M]$. For any $i\in\{1,2\}$, the sizes of $\bar{\mathbf{X}}_{ia}(t)$, $ \bar{\mathbf{X}}_{ib}(t)$, $\bar{\mathbf{Y}}_{ic}(t)$ and $\bar{\mathbf{Y}}_{id}(t)$  are the same as those of $\mathbf{X}'_{ia}(t)$, $\mathbf{X}'_{ib}(t)$, ${\mathbf{Y}}'_{ic}(t)$ and $
{\mathbf{Y}}'_{id}(t)$ respectively. {\color{black}Note that for any $i\in\{1,2\}$ and $t\in[n]$, the coefficients in linear combinations $L_{i1}(t)$ and $L_{i2}(t)$ are arbitrary 
realizations of channels, for which we allow perfect CSIT (does not hurt the outer bound argument). However, since these are realizations of channels they must satisfy all assumptions that channels are required to satisfy, e.g., $D(t)\ge\Delta_1$ where $D(t)$ is defined in (\ref{deter}) and  the fact that channel coefficients are bounded away from zero. Note that the transmitted symbols are allowed to depend on the realizations of the channel coefficients that appear in $L_{ij}$ terms since these channel coefficients are known to the transmitters. However, the realizations of the channel coefficients that appear in the $L^b_{ij}$ terms are not known to the transmitters. For these channel coefficients, the transmitted symbols can only depend on their (bounded) probability density functions, but must be independent of the actual realizations.

\subsection{A Key Lemma} 
The essential challenge in interference channels is that information sent to one receiver causes interference at the other receiver. Bounding the difference of these two terms in the GDoF sense is the key to obtaining tight GDoF outer bounds. Suppose we only wish to send information  to receiver $2$, while limiting interference at receiver $1$ as much as possible. As the first scenario, suppose we silence transmitter $1$ entirely. Then how much larger could the entropy of the signal seen at receiver $2$ be made relative to the entropy of the signal at receiver $1$? Furthermore, to strengthen the bound,  consider a second scenario where transmitter $1$ is also allowed to participate (cooperatively with transmitter $2$) but in a way that it can only be heard by receiver $2$, and not by receiver $1$. How large can the difference of entropies be made in this case? The following lemma answers these two questions, which end up being useful to derive the tight GDoF outer bounds needed for Theorem \ref{theorem:GDoF}. Note that $\bar{\bf U}$ and $\bar{\bf U}'$ stand for the effective received signals at receivers $1$ and $2$ respectively, and the two scenarios mentioned above correspond to $\gamma=0$ and $\gamma=\alpha$, respectively.

\begin{lemma}\label{lemma1} Define the two random variables $\bar{\bf U}$ and  $\bar{\bf U}'$ as,
\begin{align}
\bar{\bf U}&=\left({U}_{1}^{[n]},{U}_{2}^{[n]},\cdots,{U}_{N}^{[n]}\right)\\
\bar{\bf U}'&=\left({U'}_{1}^{[n]},{U'}_{2}^{[n]},\cdots,{U'}_{N}^{[n]}\right)
\end{align}
where for any $j\in[N]$ and $t\in[n]$ we define,
\begin{align}
U_j(t)=&L_{j1}(t)\left((\bar{\mathbf{X}}_{2a}(t))^{\alpha}\right)+L_j^b(t)\left((\bar{\mathbf{X}}_{2b}(t))^{\alpha-\beta}\right)\label{x22}\\
U'_j(t)=&\left\{
\begin{array}{ll} 
L_{j2}(t)\left({(\bar{\mathbf{X}}_{2a}(t))}^1;{(\bar{\mathbf{X}}_{2b}(t))}^1;(\bar{\mathbf{X}}_{1a}(t))^{\gamma};(\bar{\mathbf{X}}_{1b}(t))^{(\gamma-\beta)^+}\right),&1\leq j\leq \hat{N}\\ 
L_{j3}(t)\left({(\bar{\mathbf{X}}_{2a}(t))}^1;(\bar{\mathbf{X}}_{1a}(t))^{\gamma};(\bar{\mathbf{X}}_{1b}(t))^{(\gamma-\beta)^+}\right), &\hat{N}<j\le N.
\end{array}
\right.\label{x2}
\end{align}
$\gamma$ is an arbitrary positive number not greater than one. Further, let $\mathcal{W}$ be independent of $\mathcal{G}$. Then, we have,
\begin{align}
&H({\bar{\bf U}}'\mid \mathcal{W},\mathcal{G})-H({\bar{\bf U}}\mid \mathcal{W},\mathcal{G})\nonumber\\
&\le \hat{N}\max(1-\alpha+\beta,\gamma)n\log{\bar{P}}+(N-\hat{N})\max(1-\alpha,\gamma)n\log{\bar{P}}+n~o~(\log{\bar{P}})\label{lemma1q}
\end{align}
\end{lemma}
For proof of Lemma \ref{lemma1}, see Appendix \ref{app1}. The proof relies on the aligned image sets (AIS) approach of \cite{Arash_Jafar_PN}, and involves rather non-trivial generalizations because of the combination of multiple receive antennas and partial CSIT. For example, note that of the $N$ spatial dimensions  in $U_j(t)$, only $N-M$ see bounded density linear combination terms, i.e., $\bar{\mathbf X}_{2b}(t)$,  while all $N$ see the arbitrary linear combination terms $\bar{\mathbf X}_{2a}(t)$.

\subsection{Intuitive understanding of  Lemma \ref{lemma1}}
\begin{figure}[!h]
\centerline{\includegraphics[width=6in]{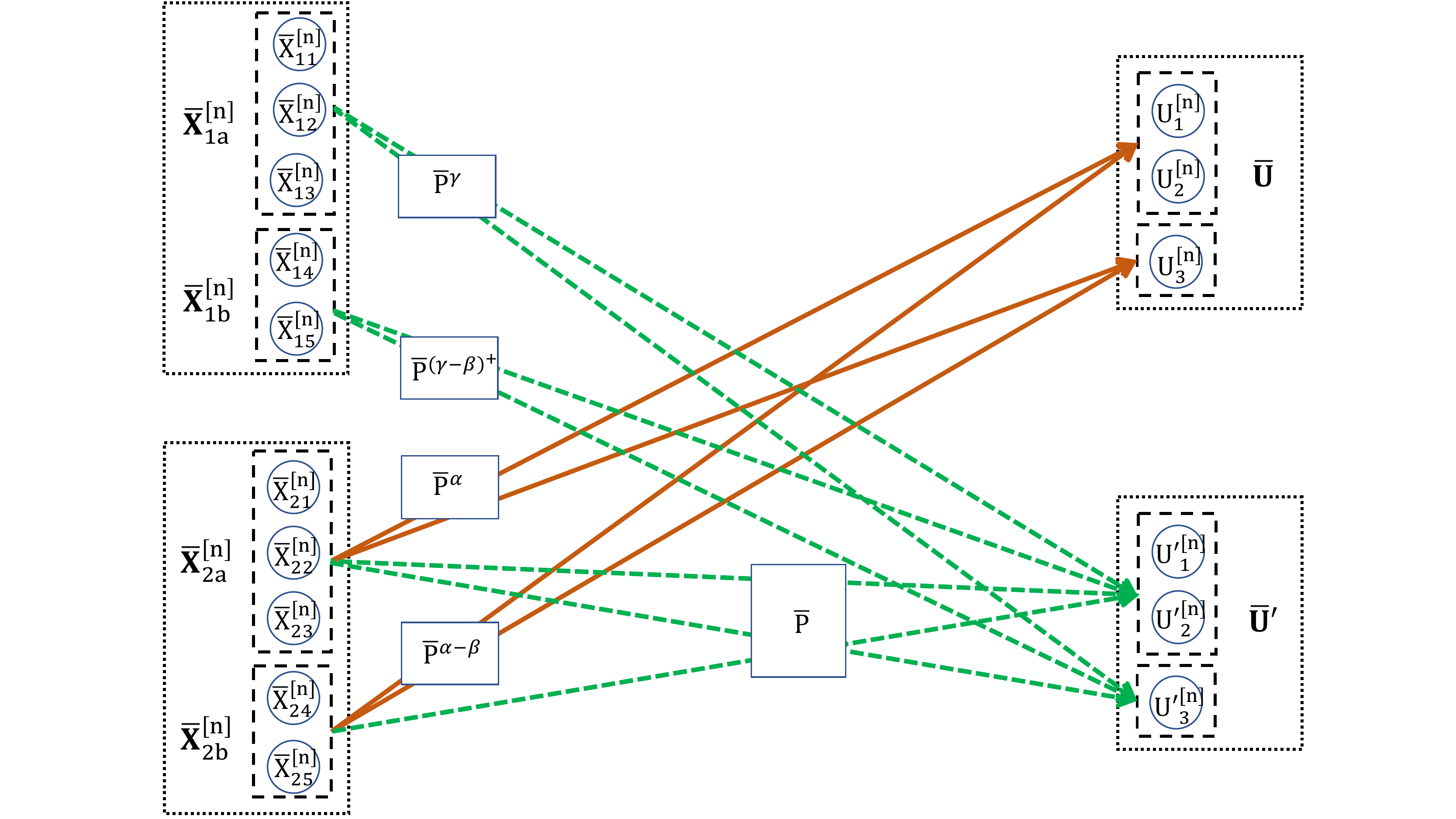}}
\caption{Specialization for Lemma \ref{lemma1} - Two User $5\times3$ MIMO IC. The channel strength terms, e.g., $\bar{P}^{\alpha-\beta}$ represent the maximum \emph{received} signal strength that can be delivered by the corresponding channels.}\label{fig:intuit}
\end{figure}
Let us use the two user $5\times3$ MIMO IC setting to provide an intuitive understanding of Lemma \ref{lemma1}. Consider inequality (\ref{lemma1q}). The left hand side of it is the difference of entropies $H({\bar{\bf U}}'\mid \mathcal{W},\mathcal{G})-H({\bar{\bf U}}\mid \mathcal{W},\mathcal{G})$, i.e., the difference of entropies of signals seen by the two receivers as illustrated in Figure \ref{fig:intuit}. We also suppress the time-index $t$ in this section to simplify the notation.  

Consider the first $\hat{N}=2$ antennas at each of the two receivers, i.e., $({U'}_1,{U'}_2)$ versus $({U}_1,{U}_2)$. Based on the channel strengths, the inputs in $\bar{X}_{2a}$ are capable of delivering $\hat{N}$ GDoF to the signals $({U}'_1,{U}'_2)$ seen by receiver $2$ while they contribute only $\hat{N}\alpha$ GDoF to $({U}_1,{U}_2)$ at receiver $1$. Thus, these inputs can contribute a difference of entropies at most equal to $\hat{N}(1-\alpha)^+$ GDoF. Similarly, the inputs $\bar{X}_{2b}$ are capable of delivering $\hat{N}$ GDoF to $({U}'_1,{U}'_2)$ seen by receiver $2$ while they contribute only $\hat{N}(\alpha-\beta)$ GDoF to $({U}_1,{U}_2)$ at receiver $1$. Thus, these inputs can at most contribute a difference of entropies equal to $\hat{N}(1-\alpha+\beta)^+$ GDoF. Similarly, the inputs $\bar{X}_{1a}$ can contribute a difference of entropies 
at most equal to $\hat{N}\gamma$ and the inputs $\bar{X}_{1b}$ can contribute a difference of entropies at most equal to $\hat{N}(\gamma-\beta)^+$. Taking the maximum across all these possibilities, the difference of entropies that can be created between 
$({U}'_1,{U}'_2)$ and $({U}_1,{U}_2)$ is at most $\hat{N}\max(1-\alpha+\beta,\gamma)$ in the GDoF sense.

Now consider the remaining $N-\hat{N}=1$ antenna at each receiver, i.e., ${U}'_3$ versus ${U}_3$. Based on channel strengths, the input $\bar{X}_{2a}$ can contribute a difference of entropies that is at most $(N-\hat{N})(1-\alpha)^+$ GDoF, $\bar{X}_{2b}$ at most $0$ GDoF (because $\bar{X}_{2b}$ is not heard by receiver $2$), $\bar{X}_{1a}$ at most $(N-\hat{N})\gamma$ GDoF and $\bar{X}_{1b}$ at most $(N-\hat{N})(\gamma-\beta)^+$ GDoF. Taking the maximum across all inputs, the difference of entropies that can be created between 
${U}'_3$ and ${U}_3$ is at most $(N-\hat{N})\max(1-\alpha,\gamma)$ GDoF.

Finally, jointly considering all the $N$ antennas at each receiver across all $n$ channel uses, we add the contributions from the first $\hat{N}$ antennas and the remaining $(N-\hat{N})$ antennas, so that the  difference of entropies $H({\bar{\bf U}}'\mid \mathcal{W},\mathcal{G})-H({\bar{\bf U}}\mid \mathcal{W},\mathcal{G})$ is at most $\hat{N}\max(1-\alpha+\beta,\gamma)+(N-\hat{N})\max(1-\alpha,\gamma)$ in the GDoF sense.  This is the intuitive understanding of the  statement of Lemma \ref{lemma1}.

\subsection{Deriving the Outer Bounds}
With the aid of Lemma \ref{lemma1}, we are now ready to derive the required outer bounds for Theorem \ref{theorem:GDoF}. In particular we will  derive bounds for the two intervals of $\alpha\le1$ and $\alpha\geq\frac{2}{3}$ separately. All the outer bounds needed for Theorem \ref{theorem:GDoF} will be recovered by combining these two cases.
\subsubsection{The case $\alpha\le 1$} Starting from Fano's inequality and omitting throughout terms that are of the order $n o(\log(P))$ and thus inconsequential for GDoF, we have,
\begin{align}
&n(R_1+R_2)\nonumber\\
&\le H(\bar{\mathbf{Y}}_{1}^{[n]}\mid\mathcal{G})-H(\bar{\mathbf{Y}}_{2}^{[n]}\mid \bar{\mathbf{X}}_{2}^{[n]},\mathcal{G})+H(\bar{\mathbf{Y}}_{2}^{[n]}\mid\mathcal{G})-H(\bar{\mathbf{Y}}_{1}^{[n]}\mid \bar{\mathbf{X}}_{1}^{[n]},\mathcal{G})\label{x1}
\end{align}
Now the term $H(\bar{\mathbf{Y}}_{2}^{[n]}\mid\mathcal{G})-H(\bar{\mathbf{Y}}_{1}^{[n]}\mid \bar{\mathbf{X}}_{1}^{[n]},\mathcal{G})$ is bounded as,
\begin{align}
&H(\bar{\mathbf{Y}}_{2}^{[n]}\mid\mathcal{G})-H(\bar{\mathbf{Y}}_{1}^{[n]}\mid \bar{\mathbf{X}}_{1}^{[n]},\mathcal{G})\nonumber\\
=&H(\bar{\mathbf{Y}}_{2}^{[n]}\mid\mathcal{G})-H(\bar{\mathbf{Y}}_{1r}^{[n]}\mid\mathcal{G})\label{ree1}
\end{align}
where $\bar{\mathbf{Y}}_{1r}(t)$ is the signal seen by receiver $1$ after the contribution from transmitter $1$ is eliminated, defined as,
\begin{align}
\bar{\mathbf{Y}}_{1r}(t)=&L_{i1}(t)\left((\bar{\mathbf{X}}_{2a}(t))^{\alpha}\right)+L_i^b(t)\left((\bar{\mathbf{X}}_{2b}(t))^{\alpha-\beta}\right)\end{align}
From Lemma \ref{lemma1}, substituting $\gamma=\alpha$ we conclude that,
\begin{align}
&H(\bar{\mathbf{Y}}_{2}^{[n]}\mid\mathcal{G})-H(\bar{\mathbf{Y}}_{1r}^{[n]}\mid\mathcal{G})\nonumber\\
&\le \Big(\hat{N}\max(1-\alpha+\beta,\alpha)+(N-\hat{N})\max(1-\alpha,\alpha)\Big)n\log{\bar{P}}\label{xec1}
\end{align}
By symmetry $H(\bar{\mathbf{Y}}_{1}^{[n]}\mid\mathcal{G})-H(\bar{\mathbf{Y}}_{2}^{[n]}\mid\bar{\mathbf{X}}_{2}^{[n]},\mathcal{G})$ is bounded similarly. Applying the GDoF limit we have,
\begin{eqnarray}
d_1+d_2&\leq&2\Big(\hat{N}\max(1-\alpha+\beta,\alpha)+(N-\hat{N})\max(1-\alpha,\alpha)\Big).
\end{eqnarray}
Equivalently,
\begin{align}
d_1+d_2&\le \left\{
\begin{array}{ll} 
2N(1-\alpha)+2\hat{N}\beta,&0\leq\alpha\leq \frac{1}{2}\vspace{0.03in}\\ 
2N\alpha+2\hat{N}(1-2\alpha+\beta)^+, &\frac{1}{2}<\alpha\leq1.\vspace{0.03in}
\end{array}
\right.\label{B1}
\end{align} 

\subsubsection{The case $\alpha\geq\frac{2}{3}$} 
Starting from Fano's inequality and omitting throughout terms that are of the order $n o(\log(P))$ and thus inconsequential for GDoF, we have,
\begin{align}
n(R_1+R_2)&\le I(\bar{\mathbf{X}}_1^{[n]};\bar{\mathbf{Y}}_1^{[n]}|\mathcal{G})+I(\bar{\mathbf{X}}_2^{[n]};\bar{\mathbf{Y}}_2^{[n]}|\bar{\mathbf{X}}_1^{[n]},\mathcal{G})\\
&\leq H(\bar{\mathbf{Y}}_1^{[n]}|\mathcal{G})+H(\bar{\mathbf{Y}}_2^{[n]}|\bar{\mathbf{X}}_1^{[n]},\mathcal{G})-H(\bar{\mathbf{Y}}_1^{[n]}|\bar{\mathbf{X}}_1^{[n]},\mathcal{G})\\
&\le N\max(1,\alpha)n\log(\bar{P})+H(\bar{\mathbf{Y}}_2^{[n]}|\bar{\mathbf{X}}_1^{[n]},\mathcal{G})-H(\bar{\mathbf{Y}}_1^{[n]}|\bar{\mathbf{X}}_1^{[n]},\mathcal{G})\\
&=N\max(1,\alpha)n\log(\bar{P})+H(\bar{\mathbf{Y}}_{2r}^{[n]}\mid\mathcal{G})-H(\bar{\mathbf{Y}}_{1r}^{[n]}\mid\mathcal{G})\label{ree1}
\end{align}
where for any $i\in\{1,2\}$, $\bar{\mathbf{Y}}_{1r}(t)$ and $\bar{\mathbf{Y}}_{2r}(t)$ are defined the same as $U_i(t)$ and $U'_i(t)$ in Lemma \ref{lemma1} with $\gamma=0$. Thus, from the statement of Lemma \ref{lemma1}  we have,
\begin{align}
&H(\bar{\mathbf{Y}}_{2r}^{[n]}\mid\mathcal{G})-H(\bar{\mathbf{Y}}_{1r}^{[n]}\mid\mathcal{G})\nonumber\\
&\le \Big(\hat{N}(1-\alpha+\beta)^++(N-\hat{N})(1-\alpha)^+\Big)n\log{\bar{P}}
\end{align}
Substituting into (\ref{ree1}) and applying the GDoF limit we obtain,
\begin{eqnarray}
d_1+d_2&\leq&N\max(1,\alpha) + \Big(\hat{N}(1-\alpha+\beta)^++(N-\hat{N})(1-\alpha)^+\Big)
\end{eqnarray}
Equivalently,
\begin{align}
&d_1+d_2\leq \left\{
\begin{array}{ll} 
2N-N\alpha+\hat{N}\beta,&\frac{2}{3}<\alpha\leq 1\\
\min(2N,N\alpha+\hat{N}(1-\alpha+\beta)^+),& 1<\alpha\leq 2\\
2N, &2<\alpha
\end{array}
\right.\label {B2}
\end{align}
Note that $2N$ is the trivial upper bound for the two user MIMO IC with $N$ antennas at receivers. Combining (\ref{B1}) and (\ref{B2}), the proof of outer bound  for Theorem \ref{theorem:GDoF} is complete.

\section{Proof of Theorem \ref{theorem:GDoF}: Achievability}
\subsection{A Useful Lemma}
Consider a $(M_1+M_2)$-user multiple access channel (MAC) where each transmitter is equipped with a single antenna, the receiver has $N$ antennas, $N< M_1+M_2$, and the $N\times 1$ received signal vector ${\bf Q}$ is represented as,
\begin{align}
{\bf Q}=&\sqrt{P}\sum_{k=1}^{M_1} {\bf H}_k{ T}_k+\sqrt{P^{\alpha}}\sum_{k=M_1+1}^{M_1+M_2} {\bf H}_k{ T}_k+\sum_{m=1}^{N} \sqrt{P^{\alpha_m}}{\bf G}_mZ_m\label{mac0}
\end{align}
where  $T_1, T_2, \cdots, T_{M_1+M_2}$ are the transmitted signals, and $Z_m$ are i.i.d. Gaussian zero mean unit variance noise terms. The ${\bf H}_k, {\bf G}_m$ are $N\times 1$ generic vectors, i.e.,  generated from continuous distributions with bounded density, so that any $N$ of them are linearly independent almost surely. The transmit power constraint is  expressed as,
\begin {eqnarray}
\mbox{E}{|T_{k}|}^2&\leq&P^{-\eta_k},~\forall k\in[M_1+M_2]\label{mac1}
\end{eqnarray}
where for any $k\in[M_1+M_2]$,  $\eta_k$ is a non-negative integer. Further, define $\gamma_k$ for $k\in[M_1+M_2]$ as,
\begin {eqnarray}
\gamma_k&=&\left\{
\begin{array}{ll} 
{(1-\eta_k)}^+,&k\in[M_1]\\ 
{(\alpha-\eta_k)}^+, &\text{Otherwise}
\end{array}
\right.
\end{eqnarray}
Thus $\gamma_k$ is the received power level of user $k$ in the GDoF sense.

The GDoF region $\mathcal{D}'$ is defined as
\begin{align}
\mathcal{D}'\triangleq&\{(d'_1,d'_2,\cdots,d'_{M_1+M_2}): \exists (R'_1(P),R'_2(P),\cdots,R'_{M_1+M_2}(P))\in\mathcal{C}'(P),\nonumber\\
& \mbox{ s.t. } d'_k=\lim_{P\rightarrow\infty}\frac{R'_k(P)}{\frac{1}{2}\log{(P)}}, \forall k\in[M_1+M_2]\} \label {region}
\end{align}
 where $\mathcal{C}'(P)$ is the capacity region of the MAC described in (\ref{mac0}).

\begin{lemma}\label{lemma:mac} The GDoF tuple $(d'_1, d'_2, \cdots, d'_{M_1+M_2})$ is achievable in the multiple access channel described above if $\forall k\in[M_1+M_2]$, and $\forall S\subset[M_1+M_2]$ where $|S|=k$,
\begin{align}
\sum_{i\in S} d'_i
&\le\max_{S_2\in S,|S_2|=\min(k,N)}\sum_{i\in S_2} \gamma_i -\min_{S_1\in[N],|S_1|=\min(k,N)}\sum_{i\in S_1} \alpha_i \label{mac3}
\end{align}
\end{lemma} 
Proof of Lemma \ref{lemma:mac} is relegated to Appendix \ref{app2}.

\subsection{Proof of Achievability in Theorem \ref{theorem:GDoF}} \label{app3}
Now, let us achieve the bound (\ref{GDOFM>N}). We will suppress the time-index $t$ in this section to simplify the notation. For any $i\in\{1,2\}$ user $i$'s message $W_i$ is split into messages $(W_{ic},W_{iz},W_{in})$, representing common message, zero-forced message, and private message, respectively. The common messages $W_{ic}$ are decoded by both receivers and are encoded into the symbols $X_{i1c},X_{i2c},\cdots,X_{i{N}c}$. These codewords are transmitted through $N$ antennas along  $M\times 1$ generic unit  vectors ${\bf V}_{i1},{\bf V}_{i2},\cdots,{\bf V}_{i{N}}$. For any $i\in\{1,2\}$, $W_{iz}$ is the sub-message to be decoded by user $i$ and  zero-forced (to the extent possible with partial CSIT) for user $\bar{i}$. $W_{iz}$ is encoded to $X_{i1z},X_{i2z},\cdots,X_{i\hat{N}z}$ and is transmitted through $\hat{N}$ antennas along the $M\times 1$  generic unit vectors ${\bf V}'_{i1},{\bf V}'_{i2},\cdots,{\bf V}'_{i\hat{N}}$ within the null space of $\hat{G}_{\bar{i}i}$, i.e.,
\begin{eqnarray}
\hat{\mathbf{G}}_{\bar{i}i}&\begin{bmatrix}\mathbf{V}'_{i1}&\mathbf{V}'_{i2}&\cdots&\mathbf{V}'_{i\hat{N}}\end{bmatrix}=\mathbf{O}_{N\times\hat{N}} \label{V'V}
\end{eqnarray}
where $\mathbf{O}_{N\times\hat{N}}$ is $N\times\hat{N}$ zero matrix. Finally, for any $i\in\{1,2\}$, $W_{in}$  acts as private message to be decoded only by receiver $i$, which is below noise floor for user $\bar{i}$. $W_{in}$ is encoded to $X_{i1n},X_{i2n},\cdots,X_{i{N}n}$ and is transmitted through $N$ antennas along $N$ generic unit vectors ${\bf V}''_{i1},{\bf V}''_{i2},\cdots,{\bf V}''_{i{N}}$. The codewords $X_{i{j}n}$ carry $1-\alpha$ GDoF each for any $j\in[N]$. The transmitted and received signals are,
\begin{align}
\mathbf{X}_{i}=&\sum_{j=1}^{{N}}\mathbf{V}_{ij}{X}_{ijc}+\sum_{j=1}^{\hat{N}}\mathbf{V}'_{ij}{X}_{ijz}+\sum_{j=1}^{{N}}\mathbf{V}''_{ij}{X}_{ijn}\\
\mathbf{Y}_{i}=&\sqrt{P}{\bf G}_{ii}\mathbf{X}_{i}+\sqrt{P^{\alpha}}{\bf \hat{G}}_{i\bar{i}}\mathbf{X}_{\bar{i}}+\sqrt{P^{\alpha-\beta}}{\bf \tilde{G}}_{i\bar{i}}\mathbf{X}_{\bar{i}}+\mathbf{\Gamma}_{i}\label{TR21a}
\end{align}

\begin{enumerate}
\item{ $\alpha\le\frac{1}{2}$}\\
Our goal here is to achieve $N(1-\alpha)+\hat{N}\beta$ GDoF per user. In this case for any $i\in\{1,2\}$, user $i$'s message $W_i$ is split to $(W_{iz},W_{in})$. $X_{ikz}$ and $X_{ijn}$  are transmitted with powers
\begin {eqnarray}
\mbox{E}{|X_{ikz}|}^2&=&P^{\beta-\alpha}\\
\mbox{E}{|X_{ijn}|}^2&=&P^{-\alpha}
\end{eqnarray}
for any $k\in[\hat{N}]$ and $j\in[N]$. The codewords $X_{ikz}$ carries $\beta$ GDoF each and remember that the codewords $X_{ijn}$ carries $1-\alpha$ GDoF each. The received signals are the same as (\ref{TR21a}), while the transmitted signals are,
\begin{align}
\mathbf{X}_{i}=&\sum_{k=1}^{\hat{N}}\mathbf{V}'_{ik}{X}_{ikz}+\sum_{j=1}^{N}\mathbf{V}''_{ij}X_{ijn}
\end{align}
for any $i\in\{1,2\}$. Using Lemma \ref{lemma:mac} we claim that each receiver, e.g., receiver $1$ can decode the desired signals as a MAC. Note that the first receiver will not see the signals from the second transmitter as the signals $X_{2kz}$ are zero-forced and $X_{2jn}$ are below noise floor. For all $m\in[N]$ set $\alpha_m=0$ and define the codewords $T_{1},\cdots,T_{N+\hat{N}}$ as 
\begin {align}
T_j=&\left\{
\begin{array}{ll} 
X_{1jz},&1\le j\le \hat{N}\\ 
X_{1(j-\hat{N})n},&\hat{N}<j\le N+\hat{N}
\end{array}
\right.
\end{align}
From the received signal in (\ref{TR21a}), $T_1,\cdots,T_{N+\hat{N}}$ are decoded by the first receiver as (\ref{mac3}) is satisfied for all $k\in[N+\hat{N}]$.

\item{$\frac{1}{2}\le\alpha\le\frac{2}{3}$}\\

In order to achieve $N\alpha+\hat{N}(1-2\alpha+\beta)^+$ GDoF per user, for any $i\in\{1,2\}$, $j\in[N]$, $k\in[\hat{N}]$, the signals $X_{ijc}$ and $X_{ikz}$ are carrying $2\alpha-1-\frac{\hat{N}\min(2\alpha-1,\beta)}{N}$ and $\beta$ GDoF respectively. The independent Gaussian codebooks are sent with powers,
\begin {eqnarray}
&&\mbox{E}{|X_{ijc}|}^2=1-P^{\beta-\alpha}-P^{-\alpha}\label{fgg2}\\
&&\mbox{E}{|X_{ikz}|}^2=P^{\beta-\alpha}\\
&&\mbox{E}{|X_{ijn}|}^2=P^{-\alpha}\label{fg2}
\end{eqnarray}
for any $i\in\{1,2\}$, $j\in[N]$ and $k\in[\hat{N}]$. Using Lemma \ref{lemma:mac} we claim that each receiver, e.g., receiver $1$ can decode the desired signals as a MAC. For any $m\in[N]$ set $\alpha_m=0$ and define the codewords $T_{1},\cdots,T_{3N+\hat{N}}$ as 
\begin {align}
T_j=&\left\{
\begin{array}{ll} 
X_{1jc},&1\le j\le N\\ 
X_{1(j-N)z},&N< j\le N+\hat{N}\\ 
X_{1(j-N-\hat{N})n},&N+\hat{N}< j\le 2N+\hat{N}\\ 
X_{2(j-2N-\hat{N})c},&2N+\hat{N}< j\le 3N+\hat{N}
\end{array}
\right.\label{ffd1}
\end{align}
From (\ref{fgg2})-(\ref{fg2}), $\gamma_1,\cdots,\gamma_{3N+\hat{N}}$ are derived as,
\begin {align}
\gamma_j=&\left\{
\begin{array}{ll} 
1,&1\le j\le N\\ 
1+\beta-\alpha,&N< j\le N+\hat{N}\\ 
1-\alpha,&N+\hat{N}< j\le 2N+\hat{N}\\ 
\alpha,&2N+\hat{N}< j\le 3N+\hat{N}
\end{array}
\right.\label{ffd2}
\end{align}
From the received signal in (\ref{TR21a}), $T_1,\cdots,T_{3N+\hat{N}}$ are decoded by the first receiver as (\ref{mac3}) is satisfied for any $k\in[3N+\hat{N}]$. For instance for $k=3N+\hat{N}$, and the set $S=[3N+\hat{N}]$ we have,
\begin {align}
&2N(2\alpha-1)-2\hat{N}\min(2\alpha-1,\beta)+\hat{N}\beta+N(1-\alpha)\le N
\end{align}
which is true as $\frac{1}{2}\le\alpha\le\frac{2}{3}$.

\item{$\frac{2}{3}<\alpha\leq 1$}\\
In this case, $\min(N\alpha+\hat{N}(1-2\alpha+\beta)^+,N-\frac{N\alpha-\hat{N}\beta}{2})$ GDoF per user is achieved. Solving the inequality $2N\alpha+2\hat{N}(1-2\alpha+\beta)^+\le2N-N\alpha+\hat{N}\beta$ leads us to define $S_{e}$ and $S_{q}$ as,
\begin{align}S_e&=\{(\alpha,\beta),0\le\beta\le\alpha,\frac{2}{3}<\alpha\leq 1\}\label{jj1}\\
S_q&=\{(\alpha,\beta),0\le\beta\le\alpha,\frac{2}{3}<\alpha\leq 1,N(3\alpha-2)\le\hat{N}(2\alpha-1),\nonumber\\
&\frac{N(3\alpha-2)}{\hat{N}}\le\beta\le\frac{N(2-3\alpha)}{\hat{N}}-2+4\alpha\}\label{jj2}
\end{align}
we will achieve $N\alpha+\hat{N}(1-2\alpha+\beta)^+$ GDoF per user when $(\alpha,\beta)\in S_q$ and $N-\frac{N\alpha-\hat{N}\beta}{2}$ GDoF per user when  $(\alpha,\beta)\in S_e \cap {S_q}^{C}$. Now consider these two cases separately.

\begin{enumerate}
\item{$N\alpha+\hat{N}{(1-2\alpha+\beta)}^+$ GDoF per user is achieved when $(\alpha,\beta)\in S_q$.}\\

The encoding and decoding follow the same as the case $\frac{1}{2}\le\alpha\le\frac{2}{3}$.\\
 
\item{$N-\frac{N\alpha-\hat{N}\beta}{2}$ GDoF per user is achieved when $(\alpha,\beta)\in S_e\cap {S_q}^C$.}\\

This case follows similar to the case $\frac{1}{2}\le\alpha\le\frac{2}{3}$ with the difference that $X_{ijc}$ carries $\frac{\alpha}{2}-\frac{\hat{N}\beta}{2N}$ GDoF for any $j\in[N]$. The decoding follows the same using Lemma \ref{lemma:mac} and defining $T_{1},\cdots,T_{3N+\hat{N}}$ the same as (\ref{ffd1}). $T_1,\cdots,T_{3N+\hat{N}}$ are decoded by the first receiver as (\ref{mac3}) is satisfied for all $k\in[3N+\hat{N}]$. For instance for $k=3N+\hat{N}$, and the set $S=[3N+\hat{N}]$ we have,
\begin {align}
&2N(\frac{\alpha}{2}-\frac{\hat{N}\beta}{2N})+\hat{N}\beta+{N}(1-\alpha)\le N
\end{align}

\end{enumerate}

\item{$1<\alpha$}\\ 

In this case, $\min(N,\frac{N\alpha+\hat{N}m_{\alpha}}{2})$ GDoF per user is achieved, where $m_{\alpha}$ is defined as ${(\beta+1-\alpha)}^+$. To do so, for any $i\in\{1,2\}$, user $i$'s message $W_i$ is split to $(W_{ic},W_{iz})$. $X_{ijc}$ and $X_{ikz}$ carry $\min(\frac{\alpha}{2}-\frac{\hat{N}m_{\alpha}}{2N},1-\frac{\hat{N}m_{\alpha}}{N})$ and $m_{\alpha}$ GDoF respectively and are transmitted with powers
\begin {eqnarray}
\mbox{E}{|X_{ijc}|}^2&=&1-P^{m_{\alpha}-1}\\
\mbox{E}{|X_{ikz}|}^2&=&P^{m_{\alpha}-1}
\end{eqnarray}
for any $i\in\{1,2\}$, $j\in[N]$ and $k\in[\hat{N}]$. The transmitted signals are,
\begin{align}
\mathbf{X}_{i}=&\sum_{j=1}^{N}\mathbf{V}_{ij}X_{ijc}+\sum_{k=1}^{\hat{N}}\mathbf{V}'_{ik}{X}_{ikz}
\end{align}
while the received signals are the same as (\ref{TR21a}). Note that the vectors $\mathbf{V}_{ij}$ and $\mathbf{V}'_{ij}$ are defined in the case $\frac{1}{2}\le\alpha\le\frac{2}{3}$. Finally,
using Lemma \ref{lemma:mac} we claim that each receiver, e.g., receiver $1$ can decode the desired signals as a MAC. For any $m\in[N]$ set $\alpha_m=0$ and define the codewords $T_{1},\cdots,T_{2N+\hat{N}}$ as 
\begin {align}
T_j=&\left\{
\begin{array}{ll} 
X_{1jc},&1\le j\le N\\ 
X_{1(j-N)z},&N< j\le N+\hat{N}\\ 
X_{2(j-N-\hat{N})c},&2N< j\le 2N+\hat{N}
\end{array}
\right.
\end{align}
From (\ref{fgg2})-(\ref{fg2}), $\gamma_1,\cdots,\gamma_{2N+\hat{N}}$ are derived as,
\begin {align}
\gamma_j=&\left\{
\begin{array}{ll} 
1,&1\le j\le N\\ 
m_{\alpha},&N< j\le N+\hat{N}\\ 
\alpha,&2N< j\le 2N+\hat{N}
\end{array}
\right.
\end{align}
From the received signal in (\ref{TR21a}), $T_1,\cdots,T_{2N+\hat{N}}$ are decoded by the first receiver as (\ref{mac3}) is satisfied for all $k\in[2N+\hat{N}]$. For instance for $k=2N+\hat{N}$, and the set $S=[2N+\hat{N}]$ we have,
\begin {align}
&2N\min(\frac{\alpha}{2}-\frac{\hat{N}m_{\alpha}}{2N},1-\frac{\hat{N}m_{\alpha}}{N})+\hat{N}m_{\alpha}\le N\alpha
\end{align}
\end{enumerate}

\section{Conclusion}
In this paper, we characterized the GDoF of the two user symmetric MIMO IC with partial CSIT under the full range of  the channel strength parameter $\alpha$ and the channel uncertainty parameter $\beta$.  The technical challenge of the paper resides in the outer bound which involves non-trivial generalizations of the AIS approach to jointly account for multiple receive antennas and partial CSIT. Generalizations of this work to the GDoF region and to more than 2 users are of the immediate interest.

\section{Appendix}
\subsection{Proof of Lemma \ref{lemma1}}\label{app1}
We are only interested in the difference of entropies of ${\bar{\bf U}}'$ and ${\bar{\bf U}}$ conditioned on $\mathcal{W}$ and $\mathcal{G}$, i.e., $H({\bar{\bf U}}'\mid \mathcal{W},\mathcal{G})-H({\bar{\bf U}}\mid \mathcal{W},\mathcal{G})$. Similar to \cite{Arash_Jafar_PN} we start with functional dependence.
\subsubsection{Functional Dependence and Aligned Image Sets} From the functional dependence argument, without loss of generality ${\bar{\bf U}}$ can be made a function of ${\bar{\bf U}}',\mathcal{W},\mathcal{G}$. So, we have,\begin {eqnarray}
&&H({\bar{\bf U}}\mid \mathcal{W},\mathcal{G})+H({\bar{\bf U}}'\mid {\bar{\bf U}},\mathcal{W},\mathcal{G})\nonumber\\
&=&H({\bar{\bf U}},{\bar{\bf U}}'\mid \mathcal{W},\mathcal{G})\label{ew1}\\
&=&H({\bar{\bf U}}'\mid \mathcal{W},\mathcal{G})\label{ew2}
\end{eqnarray}
where $(\ref{ew1})$ and $(\ref{ew2})$ follow from chain rule and the fact that ${\bar{\bf U}}$ is a function of ${\bar{\bf U}}',\mathcal{W},\mathcal{G}$. For given $\mathcal{W}$ and channel realization $\mathcal{G}$, define aligned image set $\mathcal{S}_{\nu^{[n]}}(\mathcal{W}=w,\mathcal{G})$ as the set of all ${\bar{\bf U}}'$ which result in the same ${\bar{\bf U}}$. Note that ${\bar{\bf U}}$ is a function of ${\bar{\bf U}}',\mathcal{W},\mathcal{G}$. Thus, this set is defined as the set of all values of ${\bar{\bf U}}'$ which produce  the same value for ${\bar{\bf U}}$, as is produced by ${\bar{\bf U}}'={\nu^{[n]}}$. Since uniform distribution maximizes the entropy,
\begin{align}
\mathcal{D}_{\Delta}\triangleq&H({\bar{\bf U}}'\mid \mathcal{W},\mathcal{G})-H({\bar{\bf U}}\mid \mathcal{W},\mathcal{G})\nonumber\\
=& H({\bar{\bf U}}'\mid {\bar{\bf U}},\mathcal{W},\mathcal{G})\nonumber\\
\le& \max_{w}H({\bar{\bf U}}'\mid {\bar{\bf U}},\mathcal{W}=w,\mathcal{G})\label{jen00}\\
\le&\mbox{E}_{\mathcal{G}}\left\{\log{\left|\mathcal{S}_{\nu^{[n]}}(\mathcal{W}=w,\mathcal{G})\right| }\right\}\label{jen0}\\
\le&\log\left\{\mbox{E}_{\mathcal{G}}{\left|\mathcal{S}_{\nu^{[n]}}(\mathcal{W}=w,\mathcal{G})\right| }\right\}\label{jen}
\end{align}  
where  (\ref{jen00}) and (\ref{jen}) come from  independence of $W$ and $\mathcal{G}$ and the Jensen's Inequality. Now, the most crucial step is to bound the cardinality of $\mathcal{S}_{\nu^{[n]}}$ where we need to use the `Bounded Density' assumption of $\mathcal{G}$. 
\subsubsection{Bounding the Probability that Images Align}
Given ~$\mathcal{G}$, ~consider~ two ~distinct ~instances ~of ~${\bar{\bf U}}'$ ~denoted as $\lambda^{[n]}=(\lambda_1^{[n]},\lambda_2^{[n]},\cdots,\lambda_N^{[n]})$ and $\nu^{[n]}=(\nu_1^{[n]},\nu_2^{[n]},\cdots,\nu_N^{[n]})$ produced by corresponding  realizations of codewords $(\bar{\mathbf{X}}_{1}^{[n]},\bar{\mathbf{X}}_{2}^{[n]})$ denoted by $(\bar{\mathbf{E}}_{1}^{[n]},\bar{\mathbf{E}}_{2}^{[n]})$ and $(\bar{\mathbf{F}}_{1}^{[n]},\bar{\mathbf{F}}_{2}^{[n]})$, respectively. 
\begin{align}
\lambda_i(t)=&\left\{
\begin{array}{ll} 
L_{i2}(t)\left({(\bar{\mathbf{E}}_{2a}(t))}^1;{(\bar{\mathbf{E}}_{2b}(t))}^1;(\bar{\mathbf{E}}_{1a}(t))^{\gamma};(\bar{\mathbf{E}}_{1b}(t))^{(\gamma-\beta)^+}\right),&1\leq i\leq \hat{N}\vspace{0.03in}\\ 
L_{i3}(t)\left({(\bar{\mathbf{E}}_{2a}(t))}^1;(\bar{\mathbf{E}}_{1a}(t))^{\gamma};(\bar{\mathbf{E}}_{1b}(t))^{(\gamma-\beta)^+}\right), &\hat{N}<i\le N\vspace{0.03in}
\end{array}
\right.\nonumber\\
\nu_i(t)=&\left\{
\begin{array}{ll} 
L_{i2}(t)\left({(\bar{\mathbf{F}}_{2a}(t))}^1;{(\bar{\mathbf{F}}_{2b}(t))}^1;(\bar{\mathbf{F}}_{1a}(t))^{\gamma};(\bar{\mathbf{F}}_{1b}(t))^{(\gamma-\beta)^+}\right),&1\leq i\leq \hat{N}\vspace{0.03in}\\ 
L_{i3}(t)\left({(\bar{\mathbf{F}}_{2a}(t))}^1;(\bar{\mathbf{F}}_{1a}(t))^{\gamma};(\bar{\mathbf{F}}_{1b}(t))^{(\gamma-\beta)^+}\right), &\hat{N}<i\le N\vspace{0.03in}
\end{array}
\right.\label{xff2}
\end{align}
where for any $i\in\{1,2\}$ and $t\in[n]$ we define,
\begin{align}
\bar{\mathbf{E}}_{i}(t)=&[\bar{E}_{i1}(t)\ \bar{E}_{i2}(t)\ \cdots\ \bar{E}_{iM}(t)]^T\\
\bar{\mathbf{F}}_{i}(t)=&[\bar{F}_{i1}(t)\ \bar{F}_{i2}(t)\ \cdots\ \bar{F}_{iM}(t)]^T
\end{align}
From deterministic channel model in \ref{DM_1} we have $\bar{E}_{im}(t),\bar{F}_{im}(t)\in\{0, 1, \cdots, {\bar{P}}^{\max(1,\alpha)}\}$, $\forall m\in[M]$. $\mathbb{P}(\lambda^{[n]}\in \mathcal{S}_{\nu^{[n]}})$ is bounded from above in the following three steps.\\

\begin{enumerate}
\item{Bounding the probability that ${\bar{\bf U}}(\lambda^{[n]},\mathcal{W},\mathcal{G})={\bar{\bf U}}(\nu^{[n]},\mathcal{W},\mathcal{G})$.}

For any $i\in[N]$ and $t\in[n]$ we have,
\begin{eqnarray}
&&L_{i1}(t)\left((\bar{\mathbf{E}}_{2a}(t))^{\alpha}\right)+L_i^b(t)\left((\bar{\mathbf{E}}_{2b}(t))^{\alpha-\beta}\right)\nonumber\\
&=&L_{i1}(t)\left((\bar{\mathbf{F}}_{2a}(t))^{\alpha}\right)+L_i^b(t)\left((\bar{\mathbf{F}}_{2b}(t))^{\alpha-\beta}\right)\label{poi1}
\end{eqnarray}
or in the other words, for any $i\in[N]$ and $t\in[n]$ we have,
\begin{align}
 &\left|\sum_{i=1}^Nh_i\left((E_{2i}(t))^{\alpha}-(F_{2i}(t))^{\alpha}\right)\right.\nonumber\\
 &-\left.\sum_{i={N}+1}^Mg_i\left((E_{2i}(t))^{\alpha-\beta}-(F_{2i}(t))^{\alpha-\beta}\right)\right|\le M \label{poi2}
\end{align}
where  (\ref{poi2}) follows from (\ref{poi1}) as for any real number $x$, $|x-\lfloor x\rfloor|<1$. Fix the values of $i$ and $t$. For any $j\in\{{N}+1,\cdots,M\}$ and any fixed values of $g_{i}, l\in\left({N}+1,\cdots,M\right),l\neq j$ the random variable $g_j\{(E_{2j}(t))^{\alpha-\beta}-(F_{2j}(t))^{\alpha-\beta}\}$ must take values within an interval of length no more than $2{M}$.  If $(E_{2j}(t))^{\alpha-\beta}\neq(F_{2j}(t))^{\alpha-\beta}$, then $g_{j}$ must take values in an interval of length no more than $\frac{2{M}}{|(E_{2j}(t))^{\alpha-\beta}-(F_{2j}(t))^{\alpha-\beta}|}$, the probability of which is no more than $\frac{2{M}f_{\max}}{|(E_{2j}(t))^{\alpha-\beta}-(F_{2j}(t))^{\alpha-\beta}|}$. Thus, the probability of alignment is  bounded by
\begin{align}
\mathbb{P}(\lambda^{[n]}\in \mathcal{S}_{\nu^{[n]}})\le\prod_{i=1}^N \prod_{t=1,A(t)\neq0}^n\frac{2Mf_{\max}}{{A(t)}}\label{pal}
\end{align}
where $A(t)$ is defined as 
\begin{align}
A(t)=&\max_{j\in\{{N}+1,\cdots,M\}}|(E_{2j}(t))^{\alpha-\beta}-(F_{2j}(t))^{\alpha-\beta}|
\end{align}
\item{Bounding $|\lambda_i(t)-\nu_i(t)|$ in terms of $A(t)$.}

Now, considering (\ref{poi2}) as a system of linear equations with $N$ inequalities and $N$ variables of $(E_{2i}(t))^{\alpha}-(F_{2i}(t))^{\alpha}$, we obtain, 
\begin{align}
 \left|\sum_{i=1}^Nh_i\left((E_{2i}(t))^{\alpha}-(F_{2i}(t))^{\alpha}\right)\right| \le(M-N)\Delta_2A(t)+M 
\end{align}
Following the argument presented in Appendix \ref{app:cramer}, we have,
\begin{align}
\max_{j\in[N]}|(E_{2j}(t))^{\alpha}-(F_{2j}(t))^{\alpha}|\le \frac{((M-N)\Delta_2A(t)+M)N!\Delta_2^{N-1}}{\Delta_1}\label{eq:cramer}
\end{align}
where $n!$ is defined as $\prod_{i=1}^ni$. Define $A'(t)$ as,
\begin{align}
A'(t)=&\max\left(\max_{j\in[N]}|(E_{2j}(t))^{\alpha}-(F_{2j}(t))^{\alpha}|,\max_{j\in\{N+1,\cdots,M\}}|(E_{2j}(t))^{\alpha-\beta}-(F_{2j}(t))^{\alpha-\beta}|\right)
\end{align}
From (\ref{eq:cramer}), $A'(t)$ is bounded by $c_k+c_lA(t)$, i.e., $A'(t)\le c_k+c_lA(t)$, where $c_k$ and $c_l$ are positive real numbers defined as
\begin{align}
c_k&= \frac{MN!\Delta_2^{N-1}}{\Delta_1}\\
c_l&=\frac{(M-N)N!\Delta_2^{N}}{\Delta_1}
\end{align}
From (\ref{xff2}) we bound $|\lambda_i(t)-\nu_i(t)|$ in terms of $A(t)$ as follows,
\begin{align}
|\lambda_i(t)-\nu_i(t)|\le&2M+M{\bar{P}}^{\gamma}\Delta_2+A'(t)\check{P}_i\Delta_2\\
\le&\Delta+c_lA(t)\check{P}_i\Delta_2
\end{align}
where $\Delta$ and $\check{P}_i$ are defined as,
\begin{align}
\Delta=&\lfloor2M+M{\bar{P}}^{\gamma}\Delta_2+c_k\check{P}_i\Delta_2\rfloor+1\\
\check{P}_i=&
\left\{
\begin{array}{ll} 
{M\bar{P}}^{1-\alpha+\beta},&1\leq i\leq \hat{N}\\ 
{N\bar{P}}^{1-\alpha}, &\hat{N}<i\le N
\end{array}
\right.
\end{align}
\item{} $\mathbb{P}(\lambda^{[n]}\in \mathcal{S}_{\nu^{[n]}})$ is now bounded by $|\lambda_i(t)-\nu_i(t)|$ terms as,
\begin{align}
&\mathbb{P}(\lambda^{[n]}\in \mathcal{S}_{\nu^{[n]}})\nonumber\\
\leq&\left(\prod_{i=1}^{{N}}\prod_{t:t\in[n],|\lambda_i(t)-\nu_i(t)|\leq\Delta} 1\right)\times\left( \prod_{i=1}^{{N}}\prod_{t:t\in[n],|\lambda_i(t)-\nu_i(t)|>\Delta}\frac{2Mc_{l}f_{\max}\check{P}_i\Delta_2}{|\lambda_i(t)-\nu_i(t)|-\Delta}\right)\nonumber
\end{align}
\end{enumerate}
\subsubsection{Bounding the Expected Size of Aligned Image Sets.}

\begin{align}
&\mbox{E}(|\mathcal{S}_{\nu^{[n]}}|)\nonumber\\
=&\sum_{\lambda^n}\mathbb{P}\left(\lambda^n\in \mathcal{S}_{\nu^{[n]}}\right)\nonumber\\
=&\sum_{\lambda^n}\left(\prod_{i=1}^{{N}}\prod_{t:t\in[n],|\lambda_i(t)-\nu_i(t)|\leq\Delta} 1\right)\times \left(\prod_{i=1}^{{N}}\prod_{t:t\in[n],|\lambda_i(t)-\nu_i(t)|>\Delta}\frac{2Mc_{l}f_{\max}\check{P}_i\Delta_2}{|\lambda_i(t)-\nu_i(t)|-\Delta}\right)\label{x2x}\\
\leq&\prod_{i=1}^{N}\prod_{t=1}^n\left(\sum_{\lambda_i(t):|\lambda_i(t)-\nu_i(t)|\leq\Delta}1+\sum_{\lambda_i(t):|\lambda_i(t)-\nu_i(t)|>\Delta}\frac{2Mc_{l}f_{\max}\check{P}_{i}\Delta_2}{|\lambda_i(t)-\nu_i(t)|-\Delta}\right)\nonumber\\
\leq&\prod_{i=1}^{{N}}\prod_{t=1}^n\left(2\Delta+1+2Mc_{l}f_{\max}\check{P}_{i}\Delta_2(2+2\log (1+2M\Delta_2\bar{P}))\right)\label{sigm}\\
\leq&(2Mc_{l}f_{\max}\Delta_2)^{n{N}}\bar{P}^{n\hat{N}\max(1-\alpha+\beta,\gamma)+n(N-\hat{N})\max(1-\alpha,\gamma)}\times\left(\log(\bar{P})+o(\log(\bar{P}))\right)^{nN}\label{eq:aveSK}
\end{align}
where (\ref{x2x}) follows from interchange of the summation and the product.\footnote{ Note that for the arbitrary functions $f_1(x),f_2(x),\cdots,f_n(x)$ and the arbitrary sets of numbers $S_1,S_2,\cdots,S_n$ we have,
\begin{align}
&\sum_{a_1\in S_1,a_2\in S_2,\cdots,a_n\in S_n}\prod_{t=1}^nf_t(a_t)\nonumber\\
=&\sum_{a_1\in S_1}\sum_{a_2\in S_2}\cdots\sum_{a_n\in S_n}\prod_{t=1}^nf_t(a_t)\\
=&\sum_{a_1\in S_1}f_1(a_1)\times\sum_{a_2\in S_2}f_2(a_2)\times\cdots\times\sum_{a_n\in S_n}f_n(a_n)\label{ret}\\
=&\prod_{t=1}^n\sum_{a_t\in S_t}f_t(a_t)
\end{align}} {\color{black}(\ref{sigm}) is true as the partial sum of harmonic series can be bounded above by logarithmic
function, i.e., $\sum_{i=1}^n\frac{1}{i}\le1+\log{n}$.} 
Substituting (\ref{eq:aveSK}) back into (\ref{jen}) we have,
\begin{align}
\mathcal{D}_{\Delta}\triangleq&\log\left\{\mbox{E}_{\mathcal{G}}{\left|\mathcal{S}_{\nu^{[n]}}(\mathcal{W}=w,\mathcal{G})\right| }\right\}\nonumber\\
\le&\Big(\hat{N}\max(1-\alpha+\beta,\gamma)+n(N-\hat{N})\max(1-\alpha,\gamma)\Big)n\log(\bar{P})+n~o(\log(\bar{P})\label{lkj}
\end{align}  
From (\ref{lkj}), Lemma \ref{lemma1} is concluded.

\subsection{Proof of Lemma \ref{lemma:mac}.} \label{app2}
Proof of Lemma \ref{lemma:mac} is similar to the proof of Lemma 1 in  \cite{Bofeng_Jafar_spawc16}. Consider $T_i$ for $i\in[M_1+M_2]$ as zero mean i.i.d. Gaussian random variables with power constraint defined in (\ref{mac1}). A rate tuple $(R'_1,R'_2,\cdots, R'_{M_1+M_2})$ is achievable if for any $k\in[M_1+M_2]$, and any set $S\in[M_1+M_2]$ where $|S|=k$,
\begin{eqnarray}
\sum_{i\in S}R'_i&\le& I(\{T_i,\forall i\in S\};{\bf Q}\mid\{T_j,\forall j\in{S}^C\})\label{mac4}
\end{eqnarray}
where ${S}^C$ is complement of the set $S$. (\ref{mac4}) yields,
\begin{align}
\sum_{i\in S}R'_i&\le h({\bf Q}\mid\{T_j,\forall j\in S^C\})-h({\bf Q}\mid T_1,T_2,\cdots,T_{M_1+M_2}) \label{mac5}\\
&= \max_{S_2\in S,|S_2|=\min(N,k)}\sum_{i\in S_2} \gamma_i\log{\bar{P}}+\max_{S_3\in [N],|S_3|=(N-k)^+}\sum_{i\in S_3} \alpha_i\log{\bar{P}}\nonumber\\
&-\sum_{i=1}^{N}\alpha_i\log{\bar{P}}+~o(\log{\bar{P}})\label{mac6}
\end{align}
(\ref{mac6})  yields (\ref{mac3})  in the GDoF limit.

\subsection{Justification for (\ref{eq:cramer})}\label{app:cramer}
Consider $N$ variables of $\{x_1,x_2,\cdots,x_N\}$ and $N$ inequalities of,
\begin{align}
|\sum_{j\in[N]}g_{ij}x_j|\le r_{i}, \forall i\in[N]\label{g1}
\end{align} 
where $r_i$ are non-negative real numbers and $g_{ij}$ are arbitrary realizations of channels,
for which we allow perfect CSIT (does not hurt the outer bound argument). However, since these are realizations of channels they must satisfy all assumptions that channels are required to satisfy, e.g., $D(t)\ge\Delta_1$ where $D(t)$ is defined in (\ref{deter}) and  the fact that channel coefficients are bounded away from zero. The set of solutions for (\ref{g1}) is equivalent to the union of the sets of solutions for 
\begin{align}
\sum_{j\in[N]}g_{ij}x_j= s_{i}, \forall i\in[N]\label{g2}
\end{align} 
for all $s_1,s_2,\cdots,s_N$ where $|s_i|\le r_i,\forall i\in[N]$. From Cramer's rule, any of these systems of $N$ linear equations has a solution as,
\begin{align}
x_i&=\sum_{j\in[N]}{(-1)}^{i+j}s_j\times \Lambda_{ji}\label{g3}
\end{align} 
where $\Lambda_{ji}$ is defined as,
\begin{align}
 \Lambda_{ji}&=\frac{\begin{vmatrix}
 g_{11} & \cdots&g_{1(i-1)}& g_{1(i+1)}  & \cdots&g_{1N} \\
  \vdots&\vdots& \vdots&\vdots  &\ddots  &\vdots \\  
 g_{(j-1)1} & \cdots&g_{(j-1)(i-1)} &g_{(j-1)(i+1)}  & \cdots&g_{(j-1)N} \\ 
 g_{(j+1)1} & \cdots&g_{(j+1)(i-1)} &g_{(j+1)(i+1)}  & \cdots&g_{(j+1)N} \\  
 \vdots&\vdots& \vdots&\vdots  &\ddots  &\vdots \\ 
 g_{N1} & \cdots&g_{N(i-1)}&g_{N(i+1)}  & \cdots&g_{NN} 
\end{vmatrix}}{\begin{vmatrix}
 g_{11}&g_{12}  & \cdots&g_{1N} \\ 
 g_{21}&g_{22}  & \cdots&g_{2N} \\ 
 \vdots&\vdots  &\ddots  &\vdots \\ 
 g_{N1}&g_{N2}  & \cdots&g_{NN} 
\end{vmatrix}}\label{g4}
\end{align} 
Note that from the definition of $D(t)$ in (\ref{deter}), $\Delta_1\le D(t)$ and the fact that $|g_{ij}|\le\Delta_2$, for any $i,j\in[N]$, $|\Lambda_{ji}|$ is bounded by
\begin{align}
 |\Lambda_{ji}|\le&\frac{(N-1)!\Delta_2^{N-1}}{\Delta_1}\label{g5}
\end{align} 
where (\ref{g5}) is true as absolute value of determinant of any $n\times n$ matrix with elements bounded by some number $c$, i.e., absolute value of any element of the matrix is less than $c$, is bounded by $n!c^n$. From (\ref{g3}) and (\ref{g5}), $|x_i|$ is bounded as,
\begin{align}
|x_i|&\le \sum_{j\in[N]}r_j\frac{(N-1)!\Delta_2^{N-1}}{\Delta_1}\label{ineq}
\end{align}

\bibliographystyle{IEEEtran}
\bibliography{Thesis}
\end{document}